\documentclass{article}

\usepackage{graphicx} % more modern
\usepackage{subfigure}
\usepackage{url}
\usepackage{natbib}

\usepackage{algorithm}
\usepackage{algorithmic}

\usepackage{graphicx}
\usepackage{url}
\usepackage{amsmath}
\usepackage{amssymb}
\usepackage{units}
\usepackage{wrapfig}
\usepackage{epsfig}
\usepackage{multirow}

\usepackage{booktabs}

\usepackage{hyperref}

\usepackage[accepted]{icml2013}
\icmltitlerunning{Learning from Collective Intelligence in Groups}

\begin{document}
\twocolumn[
\icmltitle{Learning from Collective Intelligence in Groups} %Learning Collective Intelligence in Groups}

% It is OKAY to include author information, even for blind
% submissions: the style file will automatically remove it for you
% unless you've provided the [accepted] option to the icml2012
% package.
\icmlauthor{Guo-Jun Qi}{qi4@illinois.edu}
\icmladdress{Beckman Institute, University of Illinois at Urbana-Champaign,
             405 N. Mathews, Urbana, IL 61801}
\icmlauthor{Charu Aggarwal}{charu@us.ibm.com}
\icmladdress{IBM T.J. Watson Research Center,
            1101 Kitchawan Road, Route 134, Yorktown Heights, NY 10598}
\icmlauthor{Pierre Moulin}{moulin@ifp.uiuc.edu}
\icmladdress{Beckman Institute, University of Illinois at Urbana-Champaign,
             405 N. Mathews, Urbana, IL 61801}
\icmlauthor{Thomas Huang}{huang@ifp.uiuc.edu}
\icmladdress{Beckman Institute, University of Illinois at Urbana-Champaign,
             405 N. Mathews, Urbana, IL 61801}

% You may provide any keywords that you
% find helpful for describing your paper; these are used to populate
% the "keywords" metadata in the PDF but will not be shown in the document
\icmlkeywords{multi-source sensing, collective intelligence, group reliability}

\vskip 0.3in
]

\begin{abstract}
Collective intelligence, which aggregates the shared information from large crowds,
is often negatively impacted by unreliable information sources with the low quality data.
This becomes a barrier to the effective use of collective intelligence in a variety of applications.
In order to address
this issue, we propose a probabilistic model to jointly assess the
reliability of sources and find the true data.
We observe that different sources
are often not independent of each other.  Instead, sources are
prone to be mutually influenced, which makes them dependent
when sharing information with each other. High
dependency between sources makes collective intelligence
vulnerable to the overuse of redundant (and possibly incorrect)
information from the dependent sources. Thus, we reveal the latent group structure among dependent sources, and
aggregate the information at the group level rather than
from individual sources directly.
This can prevent the collective intelligence from being inappropriately dominated by dependent sources.
We will also explicitly reveal the reliability of groups, and minimize the negative impacts of unreliable groups.
Experimental results on
real-world data sets show the effectiveness of the proposed approach
with respect to existing algorithms.
\end{abstract}

\section{Introduction}
Collective intelligence aggregates contributions from
multiple sources in order to collect data for a  variety of  tasks.
For example, voluntary participants collaborate with each other to
create a fairly extensive set of  entries in {\em Wikipedia}; a
crowd of paid persons may perform image and news article annotations
in {\em Amazon Mechanical Turk}.
%In addition to the crowdsourced tasks, an array of networked sensors can be deployed
%as information sources
%to monitor the locations and status of objects intruding the secured area.
These crowdsourced tasks usually involve multiple {\em objects}, such as
Wikipedia entries and images to be annotated.
The participating sources collaborate to claim their own {\em observations},
 such as  facts and labels, on these objects.
Our goal is to aggregate these collective observations to infer the
{\em true values} (e.g., the true fact and image label) for  the
different objects
\cite{Zhao:PVLDB12,Pasternack:COLING10,Galland:WSDM10}.

We note that an important property of collective intelligence is that
different sources are
typically not independent of one another.  For example, in the same
social community, people often influence each other, where their
judgments and opinions are not independent. In addition, task
participants may obtain their data and knowledge from the same
external information source, and their contributed information will
be dependent. Thus, it  may not be advisable to treat sources
independently and directly aggregate the information from individual
sources, when the aggregation process is clearly impacted by such
dependencies. In this paper, we will infer the source dependency by revealing
latent group structures among involved sources. Dependent sources will be grouped,
and their reliability is analyzed at the group level. The
incorporation of such dependency analysis  in group structures can
reduce the risk of overusing the observations made by the dependent
sources in the same group, especially when these observations are
unreliable. This helps prevent dependent sources from
inappropriately dominating collective intelligence especially when these source
are not reliable.

Moreover, we note that groups are not equally reliable, and they may
provide incorrect observations which conflict with each other, either unintentionally or maliciously.
Thus, it is important to reveal the reliability of each group, and minimize the negative impact
of the unreliable groups.  For this purpose,
we study the {\em general} reliability of each group, as well as
its {\em specific} reliability
on each individual object. These two types of reliability are closely related.
General reliability measures the overall performance of a group by aggregating each individual
reliability over the entire set of objects.
On the other hand, although each object-specific reliability is distinct, it can be better estimated
with a prior that a {\em generally reliable} group
is likely to be reliable on an individual object and vice versa.  Such prior can reduce the overfitting risk of
estimating each object-specific reliability,
especially considering that we need to determine the true value of each object at the same time \cite{Kasneci:WSDM11,Bachrach:ICML12}.

%The proposed model will explicitly explore the relationship between the general and the object-specific
%reliability, which captures
%
%while a group can have varied reliability on different objects,
%it is reasonable to assume that
%a generally reliable group
%is likely to be reliable on an individual object and vice versa.
%We will see that such assumption sets a prior on object-specific reliability which can reduce the overfitting risk of estimating them on different objects.

%as well as study the specific reliability
%
%the groups and their member sources are not equally reliable, and
%Thus, we study the reliability
%
%We note that the assessment of source reliability and the inference of true values
%are closely interdependent.
%One one hand, sources often provide their
%individual observations, and their reliability can be assessed by evaluating the correctness of
%their observations. The sources
%with higher probability of providing  true values are more likely
%to be reliable. On the other hand, however, the true values are not available in
%most crowdsourcing tasks, and we need to infer them from the observations
%contributed by individual sources \cite{Kasneci:WSDM11} \cite{Bachrach:ICML12}.
%We can better
%infer the true values from the observations made by reliable
%sources, and minimize the negative impact of those made by the unreliable sources.
%In this paper, we will explore the interdependency between source reliability and
%true values to enhance the collective intelligence from the observations made by unequally reliable sources.

The remainder of this paper is organized as follows.  In Section 2,
we formally define our problem and notations in the paper.
The Multi-Source Sensing (MSS) model for the problem is developed in Section 3, followed by
the group observation models in Section 4.
Section 5 presents the inference algorithm.  Then we
evaluate the approach in Section 6 on real
data sets, and conclude the paper in Section 7.

\section{Problem and Notational Definitions}\label{Sec:MSS:Def}

\begin{figure*}[tp]
\begin{minipage}{0.5\textwidth}
\centering
\epsfig{file=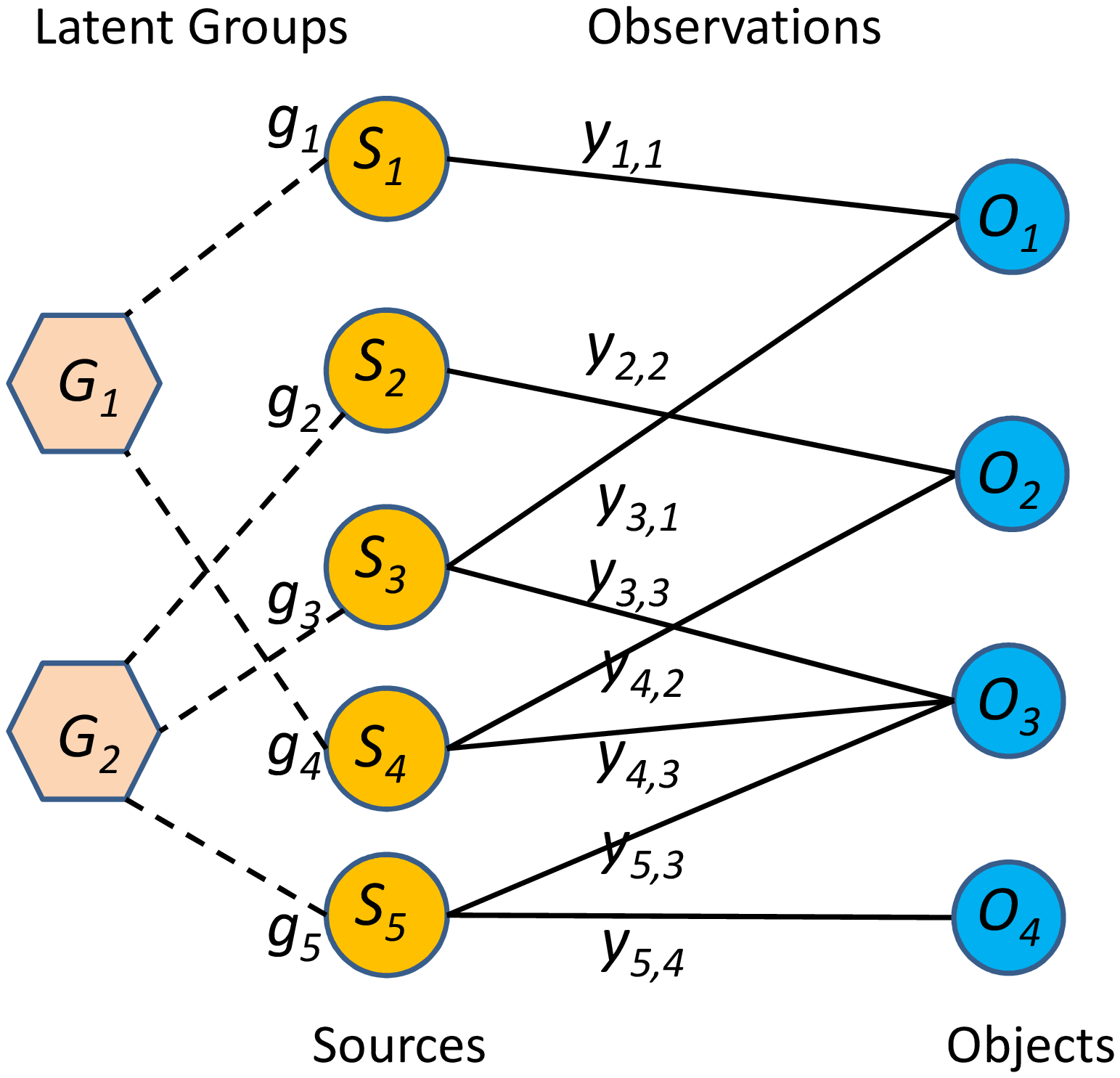,width=2.6in}
\vspace{2mm}
\caption{An example illustrating a set of five sources with their observations on four objects.}\label{Fig:Fig01}
\end{minipage}
%\hfill
%\hspace{2mm}
\begin{minipage}{0.5\textwidth}
\centering
\epsfig{file=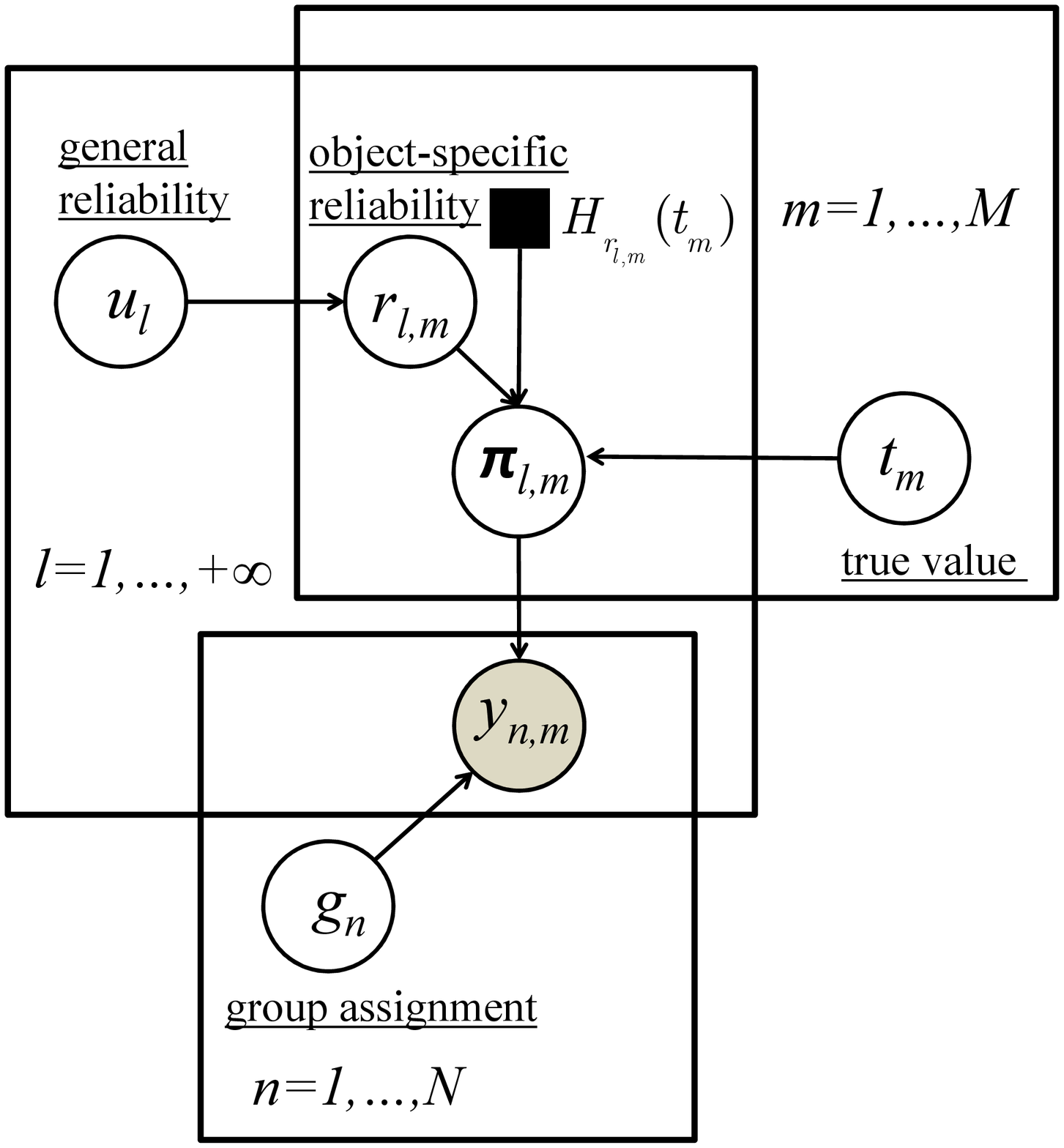,width=2.5in}
\caption{The graphical model for multi-source sensing.}\label{Fig:Fig02}
\end{minipage}
\end{figure*}

We formally define the following Multi-Source Sensing (MSS) model
which abstracts the description of collective intelligence.
Suppose that we have a set $\mathcal S:=\{S_1,S_2,\cdots,S_N\}$ of $N$ sources, and a set
$\mathcal O := \{O_1,O_2,\cdots,O_M\}$ of $M$ objects.
Each object $O_m$ takes a value $t_m$ from a domain $\mathcal X_m$ which
describes one of its attributes.  Each source $S_n$ in $\mathcal S$ reports its observation $y_{n,m}\in \mathcal X_m$ on an object $O_m$.
Then the goal of the MSS model is to infer
the true value $t_m$ of each object $O_m$ from the observations made by sources.

In this paper, we are particularly
interested in categorical domain $\mathcal X_m=\{1,\cdots,K_m\}$ with discrete values.  For example, in many crowdsourcing
applications, we focus on the (binary-valued) assertion correctness in hypothesis test
and (multi-valued) categories in classification problem.  However, the MSS model
can be straightly extended to continuous domain.  Due to the space limitation, we
leave this topic in the extended version of this paper.

Figure \ref{Fig:Fig01} illustrates an example, where five sources
make their observations on four objects. An object can be an image
or a biological molecule, and an annotator or a biochemical expert
(as a source) may claim the category (as the value) for each object.
Alternatively, an object can be a book, and a book seller web site
(as a source) claims the identity of its authors (as the values).
%Alternatively, an object may be a city, and a contributor  in
%Wikipedia (source) claims its population (value).
In a broader sense, objects are even not concrete objects.
They can refer to any crowdsourced tasks, such as
questions (e.g., ``is Peter a musician?") and
assertions (e.g., ``George Washington was born on February 22, 1732." and
``an animal is present in an image,"), and the observations
by sources are the answers to the questions,
or binary-valued positive or negative claims on these assertions.

It is worth noting that each source does not need to claim the
observations on all objects in $\mathcal O$.  In many
tasks, sources make claims only on small subsets of
objects of interest.
% The model only involves the observed, and the missing measurements can be skipped.
Thus, for notational convenience, we denote all claimed observations
by $\mathbf y$ in bold, and use $ I=\{(n,m)|\exists~y_{n,m}\in
\mathbf y\}$ to denote all the indices in $\mathbf y$. We use the
notations  $ I_{n,\cdot}=\{m|\exists~(n,m)\in I\}$ and
$I_{\cdot, m}=\{n|\exists~(n,m)\in I\}$ to denote the subset of indices
that are consistent with the corresponding subscripts $n$ and $m$.

Meanwhile, to model the dependency among sources, we assume that
there are a set of latent groups $\{G_1, G_2, \cdots\}$, and each
source $S_n$ is assigned to one group $G_{g_n}$ where
$g_n\in\{1,2,\cdots\}$ is a random variable indicating its
membership. For example, as illustrated in Figure \ref{Fig:Fig01},
the five sources are inherently drawn from two latent groups, where
each source is linked to the corresponding group by dotted lines.
Each latent group contains a set of sources which are influenced by
each other and tend to make  similar observations on objects. The
unseen variables of group membership will be inferred mathematically
from  the underlying observations.  Here, we do not assume any prior
knowledge on the number of groups. The composition of these latent
groups will be determined with the use of a Bayesian nonparametric
approach by stick-breaking construction \cite{Sethuraman:SS94},
as to be presented in the next section.

To minimize the negative impact of unreliable groups,
we will explicitly model the
group-level reliability.
Specifically, for each group $G_l$, we define a group reliability score
$u_l\in [0,1]$ in unit interval. This value measures
the general reliability of the group over the entire set of objects.
 %It represents the probability that it is a reliable group.
The higher value of
$u_l$ indicates the greater reliability of the group.
Meanwhile, we also specify the reliability $r_{l,m}\in\{0,1\}$ of each group $G_l$
on each particular object $O_m$.  When $r_{l,m}=1$, group $G_l$ will have reliable performance on $O_m$, and otherwise
it will be unreliable.
In the next section, we will clarify the relationship between general reliability $u_l$
and object-specific reliability $r_{l,m}$.

%In the next section, we will see that this general reliability of a group can be represented
%by its individual reliability distributed over the entire set of objects $\mathcal O$.
%greater reliability for $G_l$.
%We also note that the tasks of finding the true values
%usually have different levels of difficulty on different objects \cite{Bachrach:ICML12}.
%Therefore, we define $v_m\in[0,1]$ to reveal the difficulty level of
%the task associated with each object $O_m$.

\section{Multi-Source Sensing Model}\label{Sec:MSS:GP}

In this section, we present a generative process for the
multi-source sensing problem. It defines a group reliability
structure to find the dependency between sources at the same time when
we infer their
reliability at the group level.

First we define the following generative model for multi-source
sensing (MSS) process below,  the details of which will be explained
shortly.
\begin{equation}\label{Eq:Eq01}
\begin{array}{l}
\boldsymbol\lambda \sim \text{GEM}(\kappa),~~g_n|\boldsymbol\lambda \sim \text{Discrete}(\boldsymbol\lambda),
\end{array}
\end{equation}
\begin{equation}\label{Eq:Eq02}
\begin{array}{l}
u_l\sim \text{Beta}(b_1,b_0), r_{l,m} \sim \text{Bern}(u_l),t_m \sim \text{Unif}\\
\end{array}
\end{equation}
%\begin{equation}\label{Eq:Eq02b}
%\small
%\begin{array}{l}
%v_{m}\sim \text{Beta}(c_1,c_0), d_{l,m} \sim \text{Bern}(v_m), t_m \sim \text{Unif}\\
%\end{array}
%\end{equation}
\begin{equation}\label{Eq:Eq03}
\begin{array}{l}
{\boldsymbol\pi _{l,m}}|r_{l,m},{t_{m}} = z \sim H_{r_{l,m}}(t_{m})\\
\end{array}
\end{equation}
\begin{equation}\label{Eq:Eq04}
\begin{array}{l}
y_{n,m}|\boldsymbol\pi_{l,m},g_n \sim F(\boldsymbol\pi_{g_n,m})\\
\end{array}
\end{equation}
for $n=1,2,\cdots,N, m=1,2,\cdots,M, l=1,2,\cdots$.
Figure \ref{Fig:Fig02} illustrates the  generative process in a
graphical representation. Here,
$g_n|\boldsymbol\lambda\sim\text{Discrete}(\boldsymbol\lambda)$
denotes a discrete distribution, which generates the value $g_n=i$
with probability $\lambda_i$; $\text{Beta}$, $\text{Bern}$ and $\text{Unif}$ stand for Beta, Bernoulli
and uniform distributions, respectively. We explain the detail of this generative process below.

In Eq. (\ref{Eq:Eq01}),
we adopt the stick-breaking construction $\text{GEM}(\kappa)$ (named
after Griffiths, Engen and McCloskey) with  concentration parameter
$\kappa\in\mathbb R^{+}$ to define the prior distribution of
assigning each source $S_n$ to a latent group $G_{g_n}$
\cite{Sethuraman:SS94}.   Specifically, in $\text{GEM}(\kappa)$, a
set of random variables $ \boldsymbol \rho=\{\rho_1, \rho_2, \cdots\}$ are
independently drawn from the Beta distribution
$\rho_i\sim\text{Beta}(1,\kappa)$.  They define the mixing weights
$\boldsymbol\lambda$ of the group membership component such that
$p(g_n=l|\boldsymbol \rho)=\lambda_l=\rho_l\prod_{i=1}^{l-1}{(1-\rho_i)}$.
Obviously, by the above stick-breaking process, we do not need the
prior knowledge of the number of groups.  This number will be determined
by capturing the degree of dependency between sources.
%The number of groups and
%the assignment of sources
% are inferred automatically during group reliability estimation.

Clearly, we can see that the parameter $\kappa$ in the above GEM construction plays the vital role of determining {\em a priori} the degree of dependency between sources.
Actually, according to the GEM construction,
we can verify that the probability of two sources $S_n$ and $S_m$ being assigned
to the same group is
\begin{equation}
\small
\begin{aligned}%{l}
&P(g_n=g_m) = \mathop\sum\limits_{l=1}^{+\infty}\mathop\mathbb E_{\boldsymbol\lambda}{P(g_n=l|\boldsymbol\lambda)P(g_m=l|\boldsymbol\lambda)}\\
&=\mathop\sum\limits_{l=1}^{+\infty}\mathop\mathbb E\limits_{\lambda_l}\lambda_l^2
=\mathop\sum\limits_{l=1}^{+\infty}\mathop\mathbb E_{\rho_l}{\rho_l^2}{\prod_{i=1}^{l-1}\mathop\mathbb E_{\rho_i}{(1-\rho_i)^2}}\\
&= \mathop\sum\limits_{l=1}^{+\infty} \dfrac{2}{(1+\kappa)(2+\kappa)}\left(\dfrac{\kappa}{2+\kappa}\right)^{l-1}=
\dfrac{1}{1+\kappa}
\end{aligned}
\end{equation}
We can find that when $\kappa$ is smaller,
source are
more likely to be assigned to the same group where they are dependent and share
the same observation model. This will yield
higher degree of dependency between sources.
As $\kappa$ increases, the probability that any two sources belong
to the same group will decrease. In the extreme case, as
$\kappa\rightarrow +\infty$, this probability will approach to zero. In
this case, all sources will be assigned to distinctive groups,
yielding complete independence between sources. This shows that the
model can flexibly capture the various degree of dependency between
sources by setting an appropriate value of  $\kappa$.

In Eq. (\ref{Eq:Eq02}), we define a Beta distribution $\text{Beta}(b_1,b_0)$ on the group reliability score
$u_l$, where $b_1$ and $b_0$ are the soft counts which specify whether a group is reliable or not a priori, respectively.  Then
object-specific reliability $r_{l,m}\in\{0,1\}$ is sampled from the Bernoulli distribution $\text{Bern}(u_l)$ to specify the group reliability on a particular object $O_m$.
%While $u_l$ measures the {\em general} reliability of group $G_l$ distributed over the entire set of objects,
%$r_{l,m}$ gives the {\em specific} reliability of the group on a particular object $O_m$.
We can find that the higher the general reliability $u_l$, the more likely $G_l$ is reliable on a particular object $O_m$ with $r_{l,m}$ being sampled to be $1$.
This suggests that a generally more reliable group is more likely to be reliable on a particular object.
In this sense, the general reliability
serves as a prior to reduce
the overfitting risk of estimating object-specific reliability in MSS model.

%In Eq. (\ref{Eq:Eq02b}), for each object $O_m$,
%the difficulty level $v_m$ of the associated task is sampled
%from a Beta distribution $\text{Beta}(c_1,c_0)$, and
%$c_1$ and $c_0$ are soft counts which specify the prior
%knowledge about whether
%the task is difficult or not, respectively.  An indicator variable
%$d_{l,m}\in\{0,1\}$ is sampled from $\text{Bern}(v_m)$.
%When $d_{l,m}=1$, it indicates that task associated with $O_m$ is difficult for group $G_l$;
%otherwise,
%the task is not difficult for that group.
%By analogy with the setting for group reliability,
%$d_{l,m}$ gives the { \em specific } difficulty of the task
%associated with $O_m$ for a particular group $G_l$,
%while $v_m$ summarizes the {\em general} difficulty of the task for the entire set of groups.
%The higher $v_m$ is, the more likely the associated task is difficulty for a particular group.
%Thus, if a task generally difficult a priori, it is also probably difficult for a particular group.

In Eq. (\ref{Eq:Eq02}), we adopt
uniform distribution as the prior on the true value $t_m$ of each object over its domain $\mathcal X_m$.
The uniform distribution sets an unbiased prior so that
true values will be completely determined a posteriori given observations in the model inference.
%
%
%We can see that if $\boldsymbol\mu=\boldsymbol 0$, it defines a
%trivial uniform prior distribution on the true values. This may not
%be satisfactory for cases where non-trivial prior information is
%available. This could happen in cases we might be able to
%access the feature representations for objects $\mathcal O$. For
%example, if the objects are genetic sequences or text documents, we
%can extract their feature descriptors to describe the genetic
%structure and document content. Therefore,
%we wish to impose a more informative prior that aggregates these
%features into the prior distribution.
%For this purpose, given a feature vector $ \mathbf x_m$
%for an object, the prior on $t_m$ becomes a conditional distribution on $\mathbf x_m$.
%It is much flexible to choose a distribution for this prior.
%For example, we can choose an exponential distribution
%$p(t_m|\mathbf x_m, W)$ as in Eq. (\ref{Eq:Eq12}),
%with each $\mu_k$ replaced by an inner product between
%$\mathbf x_m$ and the coefficient vector
%$\mathbf w_k$ from the parameters $W
%=\{\mathbf w_k|k\in \mathcal X\}$.
%The learning of the parameters
% will be described in Section 5.

Eq. (\ref{Eq:Eq03}) and Eq. (\ref{Eq:Eq04}) define
the generative process for the observations of each source
in its assigned group.
Specifically, given the group membership $g_n$, each source $S_n$ generates
its observation $y_{n,m}$ according to the corresponding group observation model
$F(\boldsymbol\pi_{g_n,m})$.
The $\boldsymbol\pi_{l,m}$ of this model is drawn
from the
conjugate prior $H_{r_{l,m}}(t_m)$ which depends on the true value $t_m$
and  the object-specific group reliability $r_{l,m}$.
%dependent on the true value $t_m$ as well as combination of
%group reliability $r_l$ and and task difficulty $d_m$.
In the next section, we will detail the specification of
$H_{r_{l,m}}(t_m)$ and $F(\boldsymbol\pi_{l,m})$
in categorical domain.
The models in other domain can be
obtained by adopting the corresponding distribution
with the analogous idea.

\section{Group Observation Models}

%In this section, we  discuss the choice
%of group observation distribution $F(\boldsymbol\pi_{l,m})$ and its
%conjugate distribution $H_{r_l}(t_m)$ for categorical values.

In categorical domain, for each group, we choose the multinomial distribution
$F(\boldsymbol\pi_{l,m})=\text{Mult}(\boldsymbol\pi_{l,m})$ as its observation
model to generate observations $y_{n,m}$ for its member sources. Its parameter $\boldsymbol\pi_{l,m}$
%given the true value $t_{m}$ and the group reliability $r_l$,
is generated by:
\begin{equation}\label{Eq:Eq05}
%\small
\begin{aligned}
&{\boldsymbol\pi _{l,m}}|{ r_{l,m}}, { t_{m}} = z \sim H_{r_{l,m}}(t_m) \\
&:= \text{Dir}(\underbrace {{\theta^{(r_{l,m})}}, \cdots }_{z - 1},\mathop {{\eta^{(r_{l,m})}}}\limits_{\mathop  \downarrow \limits_{z^{\tt th} {\tt~entry}} } , \cdots ,{\theta^{(r_{l,m})}})\notag
\end{aligned}
\end{equation}
where $\text{Dir}$ denotes Dirchlet distribution, and $\theta^{(r_{l,m})}$ and $\eta^{(r_{l,m})}$
are its soft counts for sampling the false and true values under different settings of
$r_{l,m}$.  Below we will explain how to set these soft counts under these settings.
%We use
%the superscript in parentheses for
%the indicators of group reliability and task difficulty.

%When $d_{l,m}=1$, the task associated with object $O_m$ is difficult
%for group $G_l$. In other words, this means the group has difficulty in accessing the
%true value of
%object $O_m$.  Thus,
%we set an equal soft count for the true and false values, i.e., $\theta^{(r_{l,m},d_{l,m})} = \eta^{(r_{l,m},d_{l,m})}$ for $d_{l,m}=1$.
%Then the true value is not distinguishable from the false ones, which will prevent the group from accessing the true value on this object.
%
%On the other hand, when $d_{l,m}=0$, the task associated with object $O_m$ is not difficult for group $G_l$.  In this case,
%we have two different settings for $\theta^{(r_{l,m},d_{l,m})}$ and $\eta^{(r_{l,m},d_{l,m})}$ dependent on the group reliability $r_{l,m}=1$.

For a reliable group $G_l$ on object $O_m$ (i.e., $r_{l,m}=1$),
it should be more likely to sample
the true value $t_{m}=z$ as its observation than sampling
any other false values. Thus, we should set a larger value for $\eta^{(r_{l,m})}$ than for
$\theta^{(r_{l,m})}$.

On the other hand, if group $G_l$ is unreliable on object $O_m$ (i.e., $r_{l,m}=0$),
we can distinguish between {\em careless}
and {\em malicious} groups, and set their parameters in different ways:

%\begin{itemize}
  {\bf I. careless group:} We define $G_l$ as a careless group,
  whose member sources randomly claim values for object $O_m$, no matter which value is true.
  In this case,
  an equal soft count is set for the true and false values, i.e., $\theta^{(r_{l,m})} = \eta^{(r_{l,m})}$.
  This will make the true value indistinguishable from the false ones,
  so that the member sources makes a random guess of the true value.
  %In this case, sources are either careless or incompetent to give the true value, or incompetent with the task on $O_m$.

  {\bf II. malicious group:} In this case, group $G_l$ contains malicious sources which intentionally
   provide misleading information about the true value of object $O_m$.  In other words,
   the group  tends to claim
   the false values for object $O_m$, and thus we should set
   a larger value for $\theta^{(r_{l,m})}$ than for $\eta^{(r_{l,m})}$.
   Such malicious group can still contribute certain information
   if we read its observations
   in a reverse manner.  Actually, by setting $\theta^{(r_{l,m})}>\eta^{(r_{l,m})}$,
   the MSS model gives the
   unclaimed observations larger
   weight (corresponding to larger value of $\theta^{(r_{l,m})}$) to be
   evaluated as the true value.
%\end{itemize}

\section{Model Inference}\label{Sec:MSS:Inf}
In this section, we present the
inference and learning processes.
The MSS model defines a joint distribution on
$\mathbf g=\{g_n\}$,
$\mathbf r=\{r_{l,m}\}$,
$\mathbf u=\{u_l\}$,
$\mathbf t=\{t_m\}$,
$\boldsymbol\pi=\{\boldsymbol\pi_{l,m}\}$ and the source observations $\mathbf y$. We wish to infer the tractable posterior
$p(\mathbf g, \mathbf r, \mathbf u, \mathbf t, \boldsymbol\pi|\mathbf y)$ with a
parametric family of variational distributions in the
factorized form:
\begin{equation}\label{Eq:Eq07}
\begin{array}{l}
q(\mathbf g, \mathbf r, \mathbf u, \mathbf t, \boldsymbol\pi) =  \prod\limits_n {q( g_n|\boldsymbol\varphi_{n})}
\prod\limits_{l,m} {q(r_{l,m}|\boldsymbol\tau_{l,m})}\\
\prod\limits_l{q(u_l|\boldsymbol\beta_{l})}\prod\limits_m {q(t_m|\boldsymbol{\boldsymbol\nu_m})} \prod\limits_{l,m} {q({\boldsymbol\pi _{l,m}|\boldsymbol\alpha_{l,m}})}\notag
\end{array}
\end{equation}
with parameters $\boldsymbol\varphi_n$, $\boldsymbol\tau_{l,m}$, $\boldsymbol\beta_l$,
$\boldsymbol\nu_m$ and $\boldsymbol\alpha_{l,m}$ for these factors. The
distribution and the parameter for each factor can be determined by
variational approach \cite{Jordan:MLJ99}.
%By using the conjugate priors in MSS model, in the above
%factorization, each factor $q( g_n|\boldsymbol\varphi_{n})$ is still a multinomial distribution with parameter vector $\boldsymbol\varphi_{n}$,
%$q( r_l|\boldsymbol\tau_{l})$ is the Bernoulli distribution with parameter vector $\boldsymbol\tau_{l}$, $q({ t_m}|\boldsymbol{\nu_m})$ is the exponential distribution with parameter vector $\boldsymbol{\nu_m}$, and $q({\boldsymbol\pi _{l,m}|\boldsymbol\alpha_{l,m}})$ is a Dirichlet distribution with $\boldsymbol\alpha_{l,m}$.  To avoid notation clutter, we may omit the parameters $\boldsymbol\varphi_{n}, \boldsymbol\nu_m, \boldsymbol\alpha_{l,m}$ occasionally in these posterior factors, but the parametric dependency shall be clear in the context.
Specifically, we aim to
maximize the lower bound of the log
likelihood $\log p(\mathbf y)$, i.e., $\mathcal L(q)=\mathop
{\mathbb E}_{q}
\ln {p(\mathbf g, \mathbf r, \mathbf u, \mathbf t, \boldsymbol\pi, \mathbf y)} -
\mathbb H({q(\mathbf g, \mathbf r, \mathbf u, \mathbf t, \boldsymbol\pi)})$ with the entropy function
$\mathbb H(\cdot)$ to obtain the optimal
factorized distribution. The lower bound can be maximized over one
factor while the others are fixed. This is an approach which is
similar to coordinate descent. All the factors are updated
sequentially over steps until convergence.  We derive the
details of the  steps for updating each factor below.
%It is known that it is equivalent to minimizing the Kullback-Leibler divergence between the above factorized distribution and $p(\mathbf g,\mathbf r,\mathbf t,\boldsymbol\pi |\mathbf y)$.
%By maximizing the lower bound $\mathcal L(q)$, we can obtain the following sequential updates of each factor in $q$.
%The derivation details are omitted here due to space limit and they can be found in Appendix A.

{\bf 1:} Update each factor $q(\boldsymbol\pi_{l,m}|\boldsymbol\alpha_{l,m})$ for the group observation parameter $\boldsymbol\pi_{l,m}$.
%by fixing all the other factors to maximize $\mathcal L(q)$.
By variational approach, we can verify that the optimal $q(\boldsymbol\pi_{l,m}|\boldsymbol\alpha_{l,m})$ has the form
\begin{equation}
%\small
\begin{aligned}
q(\boldsymbol \pi_{l,m} |\boldsymbol{ \alpha}_{l,m} )
&\propto \exp\{\mathop\mathbb  E\limits_{q(\boldsymbol r_{l,m}),q(t_m)}\ln p(\boldsymbol\pi_{l,m}|r_{l,m},t_m)\\
&+\sum\limits_{n\in I_{\cdot,m}}{\mathop\mathbb E\limits_{q(\boldsymbol g_n)}\ln p(y_{n,m}|\boldsymbol\pi_{l,m},g_n)}\}\\
&\propto \prod\limits_{k \in \mathcal X} {{\pi_{l,m;k}}^{{\alpha_{l,m;k}} - 1}}\notag
\end{aligned}
\end{equation}
It still
has Dirichlet distribution with the parameters
\begin{equation}\label{Eq:Eq08}
%\small
\begin{aligned}%{l}
\alpha _{l,m;k} &= \mathop\sum\limits_{n\in I_{\cdot,m}}q(g_n=l)\delta \left[\kern-0.15em\left[ {{y_{n,m}} = k}
 \right]\kern-0.15em\right]\\
 & + \mathop\sum_{r_{l,m}\in\{0,1\}} q({r_{l,m}})
 [ ({\eta^{(r_{l,m})}} - 1)q({t_{m}} = k) \\
&+({\theta ^{(r_{l,m})}} - 1)(1 - q({t_{m}} = k)) ]
%& + q({r_l} = 0)[ ({\eta _1} - 1)q({t_{m}} = k)
% + ({\eta _0} - 1)(1 - q({t_{m}} = k)) ] \notag\\
 + 1 \notag
\end{aligned}
\end{equation}
for each $k\in\mathcal X_m$, where $\delta\left[\kern-0.15em\left[A\right]\kern-0.15em\right]$
is  the indicator function which outputs $1$ if $A$ holds, and $0$
otherwise. Here we index the element in $\boldsymbol\alpha_{l,m}$ and $\boldsymbol\pi_{l,m}$ by $k$ after the colon.
 We will follow this notation convention to index the element in vectors in this paper.
%where $\left[\kern-0.15em\left[ \cdot
% \right]\kern-0.15em\right]$ is the indicator function which outputs $1$ when the condition is true and otherwise outputs $0$.

{\bf 2:} Update each factor $q(u_l|\boldsymbol\beta_l)$ for general group reliability $u_l$.
We have
\begin{equation}
%\small
\begin{aligned}
 \ln q(u_l|\boldsymbol\beta_l)&\propto \sum_m\mathop \mathbb E\limits_{q(r_{l,m})}\ln p(r_{l,m}|u_l) + \ln p(u_l|b_1,b_0)\\
& = (\mathop\sum\limits_m q_1(r_{l,m})+b_1-1)\ln u_l \\
& + (\mathop\sum\limits_m q_0(r_{l,m})+b_0-1)\ln (1-u_l)\notag
\end{aligned}
\end{equation}
where $q_i(r_{l,m})$ is short for
$q(r_{l,m}=i)$ for $i=0,1$, respectively.
We can find
 the posterior of $u_l$ still has
 Beta distribution as
$\text{Beta}(\boldsymbol\beta_l)$
with parameter
\[\boldsymbol\beta_l=[\mathop\sum\limits_m q_1(r_{l,m})+b_1, \mathop\sum\limits_m q_0(r_{l,m})+b_0].\]
We can find that the above updated parameter sums up
the posterior reliability $q_1(r_{l,m})$ and $q_0(r_{l,m})$ over all objects. This corresponds with
the intuition that the general reliability is the sum of the reliability on individual objects.

{\bf 3:} Update each factor $q( r_{l,m}|\boldsymbol\tau_{l,m})$ for the
object-specific reliability $r_{l,m}$ of group $G_l$ on $O_m$:
\begin{equation}%\label{Eq:Eq09}
\begin{aligned}
\ln q(r_{l,m}|\boldsymbol\tau_{l,m}) &\propto \mathop\mathbb E\limits_{q(t_m),q(\boldsymbol\pi_{l,m})}{\ln q(\boldsymbol\pi_{l,m}|r_{l,m},t_m)} \\
&+ \mathop\mathbb E\limits_{q(u_l)}{\ln q(r_{l,m}|u_l)}\\
%&\propto
%   \sum\limits_k q(t_m = k)
%  [ ({\eta^{(r_{l,m},d_{l,m})} } - 1)\mathop \mathbb E\limits_{q(\boldsymbol\pi_{l,m})} \ln \pi _{l,m;k}\\
%  &+ ({\theta^{(r_{l,m},d_{l,m})} } - 1)\sum\limits_{j \ne k}{\mathop \mathbb E\limits_{q(\boldsymbol\pi_{l,m})} \ln \pi _{l,m;j}}]\\
%  &+ r_{l,m}\mathbb E_{q(u_l)}{\ln u_l}+(1-r_{l,m})\mathbb E_{q(u_l)}{\ln (1-u_l)}\\
%  &+ d_{l,m}\mathbb E_{q(v_m)}{\ln v_m}+(1-d_{l,m})\mathbb E_{q(v_m)}{\ln (1-v_m)}\notag
\end{aligned}
\end{equation}
Thus, we have
\begin{equation}\label{Eq:Eq06}%\label{Eq:Eq09}
\begin{aligned}
&\ln q(r_{l,m}|\boldsymbol\tau_{l,m})  \\
&  \propto \sum\limits_{k\in\mathcal X_m} q(t_m = k)
  [ ({\eta^{(r_{l,m})} } - 1)\mathop \mathbb E\limits_{q(\boldsymbol\pi_{l,m})} \ln \pi _{l,m;k}\\
  &+ ({\theta^{(r_{l,m})} } - 1)\sum\limits_{j \ne k}{\mathop \mathbb E\limits_{q(\boldsymbol\pi_{l,m})} \ln \pi _{l,m;j}}]\\
  &+ r_{l,m}\mathop\mathbb E_{q(u_l)}{\ln u_l}+(1-r_{l,m})\mathop\mathbb E_{q(u_l)}{\ln (1-u_l)}
\end{aligned}
\end{equation}
for $r_{l,m}\in\{0, 1\}$, respectively.
%and
%\begin{equation}
%\begin{array}{l}
%\ln q({r_l} = 0) \propto
%  \sum\limits_{m} \sum\limits_k q({t_{m}} = k)[ ({\eta _1} - 1)\mathop \mathbb E\limits_{q(\boldsymbol\pi_{l,m})} \ln \pi _{l,m;k} \\
%  + \sum\limits_{j \ne k} {({\eta _0} - 1)\mathop \mathbb E\limits_{q(\boldsymbol\pi_{l,m})} \ln \pi _{l,m;j}}  ]
%\end{array}
%\end{equation}
Here we compute the expectation of the logarithmic Dirichlet
variable as
\[\mathop \mathbb E_{q(\boldsymbol\pi_{l,m})} \ln \pi
_{l,m;k}=\psi(\sum_i {\alpha_{l,m;i}})-\psi(\alpha_{l,m;k})\]
with
the digamma function $\psi(\cdot)$; the expectation of the logarithmic Beta variables
\[\mathbb E_{q(u_l)}{\ln u_l}=\psi(\beta_{l;1}+\beta_{l;2})-\psi(\beta_{l;1})\]
and
\[\mathbb E_{q(u_l)}{\ln (1-u_l)}=\psi(\beta_{l;1}+\beta_{l;2})-\psi(\beta_{l;2}).\]
%and $\alpha_{l,m;k}$ is the
%super parameters for the posterior Dirichlet distribution obtained
%in Step 1.
Finally, the updated values of $q({r_{l,m}})$ are normalized to be valid probabilities.

The last line of Eq. (\ref{Eq:Eq06}) reflects how the general reliability $u_l$ affects the
estimation of the object-specific reliability.  This embodies the idea that
a generally reliable group is likely to be reliable on a particular object and vice versa. This can reduce the overfitting risk
of estimating $r_{l,m}$ especially considering that $q(t_m)$ in the second line also need to be estimated simultaneously in MSS model
as in the next step.

{\bf 4:} Update each factor $q(t_m|\boldsymbol{\nu_m})$ for the true
value. We have
%We can update the factor by maximizing
%the lower bound $\mathcal L(q)$ as follows:
\begin{equation}
%\small
\begin{aligned}%{l}
&\ln q({t_m} = k|\boldsymbol{\nu_m}) \propto \ln p({t_m} = k) \\
&+ \mathop\sum\limits_{l}\mathop\sum\limits_{r_{l,m}\in\{0,1\}}q(r_{l,m})\mathop \mathbb E\limits_{q({\boldsymbol\pi _{l,m}})} \ln p({\boldsymbol\pi _{l,m}}|t_m=k,r_{l,m})\notag
\end{aligned}
\end{equation}
%This suggests that $q(t_m = k|\boldsymbol\nu_m)$ is still an exponential distribution
%$\text{Exp}(\boldsymbol{\nu_m})$ with the parameter
This suggests that
%\begin{equation}
%\begin{split}
%&{{ \nu }_{m;k}}
% = {\mu _k} + \mathop\sum_{l}\mathop\sum_{r_l,d_m}q(r_l,d_m)\{ ({\eta^{(r_l,d_m)}} - 1)\\
% &\times\mathop \mathbb E\limits_{q({\boldsymbol\pi _{l,m}})} \ln {\pi _{l,m;k}}
% + \sum\limits_{k' \ne k} {({\theta^{(r_l,d_m)}} - 1)\mathop \mathbb E\limits_{q({\boldsymbol\pi _{l,m}})} \ln {\pi _{l,m;k'}}}  \}\notag\\
%% &+ q(r_l=0)\{ ({\eta _1} - 1)\mathop \mathbb E\limits_{q({\pi _{l,m;k}})} \ln {\pi _{l,m;k}}  \\
%% &+ \sum\limits_{k' \ne k} {({\eta _0} - 1)\mathop \mathbb E\limits_{q({\pi _{l,m;k}})} \ln {\pi _{l,m;k'}}}  \}\\
%\end{split}
%\end{equation}
\begin{equation}
\begin{split}
&\ln q({t_m} = k|\boldsymbol{\nu_m})\\
 &\propto \mathop\sum_{l}\mathop\sum_{r_{l,m}}q(r_{l,m})\{ ({\eta^{(r_{l,m})}} - 1)
 \mathop \mathbb E\limits_{q({\boldsymbol\pi _{l,m}})} \ln {\pi _{l,m;k}}\\
 &+ \sum\limits_{k' \ne k} {({\theta^{(r_{l,m})}} - 1)\mathop \mathbb E\limits_{q({\boldsymbol\pi _{l,m}})} \ln {\pi _{l,m;k'}}}  \}\notag\\
% &+ q(r_l=0)\{ ({\eta _1} - 1)\mathop \mathbb E\limits_{q({\pi _{l,m;k}})} \ln {\pi _{l,m;k}}  \\
% &+ \sum\limits_{k' \ne k} {({\eta _0} - 1)\mathop \mathbb E\limits_{q({\pi _{l,m;k}})} \ln {\pi _{l,m;k'}}}  \}\\
\end{split}
\end{equation}
All $q(t_m=k), k\in\mathcal X_m$ are normalized to ensure they are validate probabilities.

{\bf 5:} Update each factor $q(g_n|\boldsymbol\varphi_n)$ for the group assignment
of each source. We can derive
%the following optimal factor:
%\[
%\small
%\begin{aligned}%{l}
%&\ln q({g_n} = l|\boldsymbol\varphi_n) \\
%&\propto \mathop \mathbb E\limits_{q(\boldsymbol \rho)} \ln p({g_n} = l|\boldsymbol \rho) +
%\sum\limits_{m \in {I_{n,\cdot}}} {\mathop \mathbb E\limits_{q({\boldsymbol\pi _{l,m}})} \ln p({y_{n,m}}|{\pi _{l,m}},{g_n} = l)} \\
%& = \mathop \mathbb E\limits_{q(\boldsymbol \rho)} \ln p({g_n} = l|\boldsymbol \rho) + \sum\limits_{m \in {I_{n,\cdot}}} {\sum\limits_{k \in \mathcal X_m} {\delta \left[\kern-0.15em\left[ {y_{n,m} = k}
% \right]\kern-0.15em\right]\mathop\mathbb E\limits_{q({\boldsymbol\pi _{l,m}})} \ln {\pi _{l,m;k}}} }
%\end{aligned}
%\]
\begin{equation}
\small
\begin{aligned}%{l}
&\ln q({g_n} = l|\boldsymbol\varphi_n) \\
&\propto \mathop \mathbb E\limits_{q(\boldsymbol \rho)} \ln p({g_n} = l|\boldsymbol \rho) +
\sum\limits_{m \in {I_{n,\cdot}}} {\mathop \mathbb E\limits_{q({\boldsymbol\pi _{l,m}})} \ln p({y_{n,m}}|{\pi _{l,m}},{g_n} = l)} \\
& = \mathop \mathbb E\limits_{q(\boldsymbol \rho)} \ln p({g_n} = l|\boldsymbol \rho) + \sum\limits_{m \in {I_{n,\cdot}}} {\mathop\mathbb E\limits_{q({\boldsymbol\pi _{l,m}})} \ln {\pi _{l,m;y_{n,m}}} }\notag
\end{aligned}
\end{equation}
This shows that $q({g_n} = l|\boldsymbol\varphi_n)$ is a multinomial distribution with its parameter as
\begin{equation}\label{Eq:Eq09}
\varphi_{n;l} = q({g_n} = l|\boldsymbol\varphi_n) = \dfrac{{\exp ({U_{n,l}})}}{{\sum\limits_{l = 1}^\infty  {\exp ({U_{n,l}})} }}
\end{equation}
where
%\[\begin{array}{l}
%{U_{n,l}} = \mathop \mathbb E\limits_{q(\boldsymbol \rho)} \ln p({g_n} = l|\boldsymbol \rho) \\
%+ \sum\limits_{m \in {I_n}} {\sum\limits_{k \in \mathcal X} {\delta \left[\kern-0.15em\left[ {y_{n,m} = k}
% \right]\kern-0.15em\right]\mathop \mathbb E\limits_{q({\boldsymbol\pi _{l,m}})} \ln {\pi _{l,m;k}}} }
% \end{array}\]
\begin{equation}
\begin{array}{l}
{U_{n,l}} = \mathop \mathbb E\limits_{q(\boldsymbol \rho)} \ln p({g_n} = l|\boldsymbol \rho)
+ \sum\limits_{m \in {I_{n,\cdot}}} { {\mathop \mathbb E\limits_{q({\boldsymbol\pi _{l,m}})} \ln {\pi _{l,m;y_{n,m}}}} }\notag
 \end{array}
\end{equation}

As in \cite{Kurihara:NIPS06}, we truncate after $L$
groups: the posterior distribution $q(\rho_i)$ after the level $L$ is
set to be its prior $p(\rho_i)$ from $\text{Beta}(1,\kappa)$; and all the
expectations $\mathop \mathbb E\limits_{q({\boldsymbol\pi _{l,m}})}
\ln {\pi _{l,m;k}}$ after $L$  are  set to:
\[\begin{aligned}
& \mathbb E_{q({\boldsymbol\pi _{l,m}})} \ln {\pi _{l,m;k}}
&=\mathop \mathbb E\limits_{q(t_m),p(r_{l,m})} \left\{~\mathop\mathbb E[\ln {\pi _{l,m;k}}|r_{l,m},t_m]~\right\}
%+p(r_l=0)\mathop \mathbb E\limits_{q(t_m)} \mathop\mathbb E\limits_{p(\pi_{l,m}|t_m,r_l=0)}[\ln {\pi _{l,m;k}}|t_m,r_l=0]
\end{aligned}
\]
with $p(r_{l,m})$ defined as in Eq.
(\ref{Eq:Eq02}) for all $l>L$, respectively. The inner conditional expectation in the
above is taken with respect to  the probability of
$\boldsymbol\pi_{l,m}$ conditional on $r_{l,m}$ and $t_m$ as defined in
(\ref{Eq:Eq03}). Similar to the family of nested Dirichlet process mixture in \cite{Kurihara:NIPS06},
this will form a family of nested
priors indexed by $L$ for the MSS model.  Thus, we can compute the
infinite sum in the denominator of Eq. (\ref{Eq:Eq09}) as:
\[\sum\limits_{l = L + 1}^\infty  {\exp ({U_{n,l}})}  = \frac{{\exp ({U_{n,L + 1}})}}{{1 - \exp (\mathop \mathbb E\limits_{\rho_i\sim \text{Beta}(1,\kappa)} \ln (1 - \rho_i))}}\]

{\bf 6:} Finally, we can find that before the truncation level $L$,
the posterior distribution $q({\rho_i}) \sim \text{Beta}({\phi
_{i,1}},{\phi _{i,2}})$  is updated as
\[{\phi _{i,1}} = 1 + \sum\limits_{n = 1}^N {q({g_n} = i)} ,~~{\phi _{i,2}} = \kappa  + \sum\limits_{n = 1}^N {\sum\limits_{j = i+1}^\infty  {q({g_n} = j)} } \]

The above steps are iterated to yield the optimal factors.
%Besides the inference, we need to learn
%the parameter $\boldsymbol\mu$ in $p(t_m|\boldsymbol\mu)$ or $W$ in
%$p(t_m|\mathbf x_m, W)$.  Here, we adopt the variational EM
%(Expectation-Maximization) algorithm.  In each iteration, the E-step
%(expectation) involves computing the  tractable posterior
%distributions as in the inference step.  Then, the maximization step will update
%$\boldsymbol\mu$ or $W$ by maximizing the expected log-likelihood
%over $q$ as follows:
%\begin{equation}\label{Eq:Eq11}
%\max\limits_{\boldsymbol\mu}\sum\limits_{m = 1}^M {{\mathbb E_{q({t_m|\boldsymbol\nu_m})}}\log p({t_m}|\boldsymbol\mu)}
%\end{equation}
%or
%\begin{equation}
%\max\limits_{W}\sum\limits_{m = 1}^M {{\mathbb E_{q({t_m|\boldsymbol\nu_m})}}\log p({t_m}|{\bf x_m}, W)}
%\end{equation}
%We can adopt any off-the-shelf optimization algorithms to solve the above
%problem.

\section{Experimental Results}\label{Sec:MSS:Exp}
In this section, we compare our approach with other existing
algorithms and demonstrate its effectiveness  for  inferring source
reliability together with  the true values of objects. The
comparison is performed on a book author data set from online book
stores, and a user tagging data set from the online image sharing
web site \url{Flickr.com}.
%{\bf Charu Note: citation/data source?}

{\bf Book author data set:} The first data set is the book author
data set prepared in \cite{Yin:KDD07}.  The data set is obtained by
crawling $1,263$ computer science books on {\em AbeBooks.com}.  For
each book, {\em AbeBooks.com} returns the book information extracted
from a set of online book stores.  This data set contains a total of
$877$ book stores (sources), and $24,364$ listings of books
(objects) and their author lists (object values) reported by these
book stores.  Note that each book has a different categorical domain
$\mathcal X$ that contains all the authors claimed by sources.
%In this data set, we do not extract any features to represent objects, and
%the model parameters in Eq (\ref{Eq:Eq12}) are set to $\mathbf 0$, which equivalently set a uniform prior for the true values.

Author names are normalized by preserving the first and last
names, and ignoring the middle name of each author.  For evaluation
purposes, the authors of $100$ books are manually collected from
the scanned book covers \cite{Yin:KDD07}.  We compare the returned
results of each model with the ground truth author lists on this
test set and report the accuracy.

We compare the proposed algorithm with the following ones: (1) the
naive Voting algorithm which counts the top voted author list for
each book as the truth; (2) {\em TruthFinder} \cite{Yin:KDD07}; (3)
{\em Accu} \cite{Dong:VLDB09b} which considers the dependency
between sources; (4) {\em 2-Estimates} as described in
\cite{Galland:WSDM10} with the highest accuracy among all the models
in \cite{Galland:WSDM10} (5) {\em MSS}, which is our proposed
algorithm.  In the experiments, we choose
the parameters $\eta^{(r_{l,m})}$ and $\theta^{(r_{l,m})}$ from $\{1.0, 2.0, 5.0, 10.0\}$
for $r_{l,m}\in\{0,1\}$ as in Section 4, $b_0, b_1$ from $\{1.0, 2.0, 4.0\}$, and
$\kappa$ from $\{1.0, 5.0, 10.0\}$. Due to the unsupervised nature of the problem, we pick
the set of parameters with the maximum observation likelihood.
%We fixed the truncation level at $80$ in variational inference
%for stick-breaking process.

\begin{table*}[tb]
\small
\centering \caption{Top-10 and bottom-10 book stores ranked by their posterior probability of belonging to a reliable group.
 We also report the accuracy of these bookstores on the test set. }
\vspace{+2mm}% The smallest error rate for each category is in bold.}
%\small
\begin{tabular}{l|l||l|l} \toprule
top-10 bookstore&accuracy&bottom-10 bookstore&accuracy \\\midrule
International Books &   1   &   textbooksNow    &   0.0476  \\
happybook &   1   &   Gunter Koppon   &   0.225   \\
eCampus.com &   0.9375  &   www.textbooksrus.com  &   0.3333  \\
COBU GmbH \& Co. KG &   0.875    &   Gunars Store    &   0.2308    \\
HTBOOK  &   1  &   Indoo.com   &   0.3846  \\
AlphaCraze.com  &   0.8462  &   Bobs Books  &   0.4615  \\
Cobain LLC   &   1   &   OPOE-ABE Books  &   0   \\
Book Lovers USA &   0.8667  &   The Book Depository &   0.3043  \\
Versandantiquariat Robert A. Mueller   & 0.8158  &   Limelight Bookshop  &   0.3896  \\
THESAINTBOOKSTORE   &   0.8214  &   textbookxdotcom &   0.4444  \\\bottomrule
\end{tabular}\label{Tb:Tb02}
\end{table*}

\begin{table*}
\centering
%\small
\caption{Comparison of different algorithms on book author and Flickr data set.  On book author data set, the algorithms are compared by their accuracies.  On Flickr data set, the algorithms are compared by their average precisions and recalls on $12$ tags.}\label{Tb:Tb01}\vspace{2mm}
\begin{tabular}{l|c||c|c} \toprule
Model&book author data set&\multicolumn{2}{c}{Flickr data set}\\\cline{2-4}
&accuracy&precision&recall\\\midrule
{\em Voting}\cite{Dong:VLDB09b}&0.71&0.8499&0.8511\\
{\em 2-Estimates}\cite{Galland:WSDM10}&0.73&0.8545&0.8602\\
{\em TruthFinder}\cite{Yin:WWW11}&0.83& 0.8637&0.8649\\
{\em Accu}\cite{Dong:VLDB09b}&0.87& 0.8731 &0.8743\\
%AccuWithSim\cite{Dong:VLDB09b}&0.89& \\
%{\em MSSNoRank} &0.81&0.8591&0.8556\\
{\em MSS} &{\bf 0.95}&{\bf 0.9176}&{\bf 0.9212}\\
\bottomrule
\end{tabular}
\end{table*}
%
%\begin{table*}[tb]
%\centering \caption{Top-10 book stores ranked by their membership in $G_1$.  We also report the accuracy of these bookstores on the test set. }\label{Tb:Tb02}\vspace{+2mm}
%% The smallest error rate for each category is in bold.}
%%\small
%\begin{tabular}{l|l} \toprule
%top-10 ranked bookstores&accuracy \\\midrule
%BetterWorld.com    &   1   \\
%Kayleighbug    &   1   \\
%A1Books    &   0.7791  \\
%COBU GmbH \& Co. KG    &   0.875   \\
%HTBOOK &   0.8889  \\
%AlphaCraze.com &   0.8462  \\
%The E  &   1   \\
%Book Lovers USA    &   0.8667  \\
%Versandantiquariat Robert A. Mueller   &   0.8158  \\
%THESAINTBOOKSTORE  &   0.8214  \\\bottomrule
%\end{tabular}
%\end{table*}

\begin{figure}[t]%{0.32\textwidth}
\begin{center}
\includegraphics[width=0.38\textwidth]{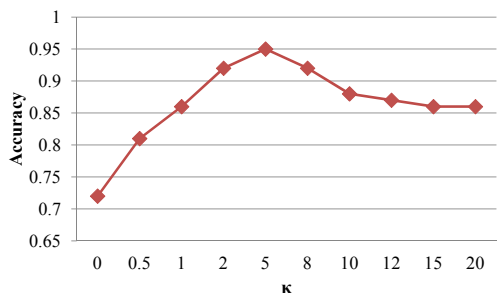}
\end{center}
\caption{Model accuracy versus different $\kappa$ on book author dataset.}\label{Fig:Fig04}
\end{figure}

 \begin{table*}[]
\centering \caption{The rounds used before convergence and computing time for each model.}% The smallest error rate for each category is in bold.}
%\small
\begin{tabular}{l|l|l||l|l} \toprule
\multirow{2}{*}{Model}&\multicolumn{2}{|c||}{Bookstore}&\multicolumn{2}{|c}{User Tagging}\\\cline{2-5}
&Rounds &Time(s) & Rounds & Time (s)\\\midrule
Voting&1&0.2& 1 & 0.5 \\
2-Estimates&29&21.2&32&628.1  \\
TruthFinder&8&11.6&11& 435.0 \\
Accu&22&185.8 & 23 & 3339.7\\
%AccuWithSim&18&197.5\\
MSS&9&10.3&12&366.2\\
\bottomrule
\end{tabular}\label{Tb:Tb03}
\end{table*}

Table \ref{Tb:Tb01} compares the results of the different algorithms
on the book author data set in terms of the accuracy. The {\em MSS}
model achieves the best accuracy among all the compared models.
%In comparison, {\em MSSNoRank} does not utilize group reliability
%ranking,  and it only achieves an accuracy of $0.74$, which is much
%lower than {\em MSS}. It shows that the process of  ranking group
%reliability plays an important role in inferring the true data
%values.
We note that the proposed {\em MSS} model is an
unsupervised algorithm which does not involve any training data.
 That is to say, we do not use any true values in the MSS algorithm
in order to produce the reliability ranking as well as other true values.
Even compared with the accuracy of $0.91$  of the
Semi-Supervised Truth Finder (SSTF) \cite{Yin:WWW11} using extra
training data with known true values on some objects, the {\em MSS }model still achieves the highest $0.95$
accuracy.  It suggests that with additional training data, the {\em
MSS} model may improve its accuracy further.

Since $\kappa$ is predicative of the dependency between sources, we
study the changes of the model accuracy versus various $\kappa$ in
Figure \ref{Fig:Fig04}. We know that when $\kappa=0$, all sources
are completely dependent, and assigned to the same group.  At this
time, the model has a much lower accuracy, since all sources are
tied to the same level of reliability within a single group.  As
$\kappa$ increases, the accuracy achieves the peak at $\kappa=5.0$.
After that, it deteriorates as the model gradually stops  capturing
the source dependency with  increased $\kappa$. This demonstrates
the importance of modeling the source dependency, and the capability of
MSS model to capture such dependency by
$\kappa$.

Moreover, to compare the reliability between sources, we
can define the reliability of each source $S_n$ by the expected reliability score
of its assigned groups as
\[
\text{Reliability}(S_n)=\sum_l q(g_n=l) \mathop\mathbb E_{q(u_l|\boldsymbol\beta_l)}[u_l]
\]
where
\[
\mathop\mathbb E_{q(u_l|\boldsymbol\beta_l)}[u_l]=\frac{\beta_{l,1}}{\beta_{l,1}+\beta_{l,2}}
\]
Then, sources can be ranked based on such source reliability.
In Table \ref{Tb:Tb02}, we rank the top-10 and bottom-10 book stores
in this way.
%Specifically,
%the book stores are ranked based on the posterior probability of
%these book stores belonging to the reliable group.
In order to show the extent to which this ranking list is consistent
with the real source reliability, we  provide  the accuracy of these
bookstores on test data sets. Note that each individual bookstore
may only claim on a subset of books in the test set, and the
accuracy is computed based on the claimed books. From the table, we
can see that the obtained rank of data sources is consistent with
the rank of their accuracies on the test set. On the contrary, the
accuracy of the bottom-10 bookstores is much worse compared to that
of the top-10 book stores on the test set. This also explains partly
the better performance of the MSS model.
%By excluding them from
%the reliable groups, the model can minimize the adverse effect of
%these low quality sources on  true
% value inference.

%It is interesting to see that although most of these top-ranked bookstores have a lower accuracy than $0.94$, by aggregating these multi-source sensing results, we achieve an overall better result.  This suggests by properly ranking the sources, we can significantly improve the

%\begin{table}
%\centering
%\caption{Comparison of accuracy of different algorithms on Flickr data set.}\label{Tb:Tb03}
%\begin{tabular}{l|l} \toprule
%Model&Accuracy\\\midrule
%VotedTag&0.71\\
%Low-Rank\cite{Qi:TPAMI12}&\\
%MSSNoRank&\\
%MSS&{\bf 0.94}\\
%\bottomrule
%\end{tabular}
%\end{table}
\begin{figure}[t]
\centering{
%\mbox{
\subfigure[balloon] {\epsfig{figure=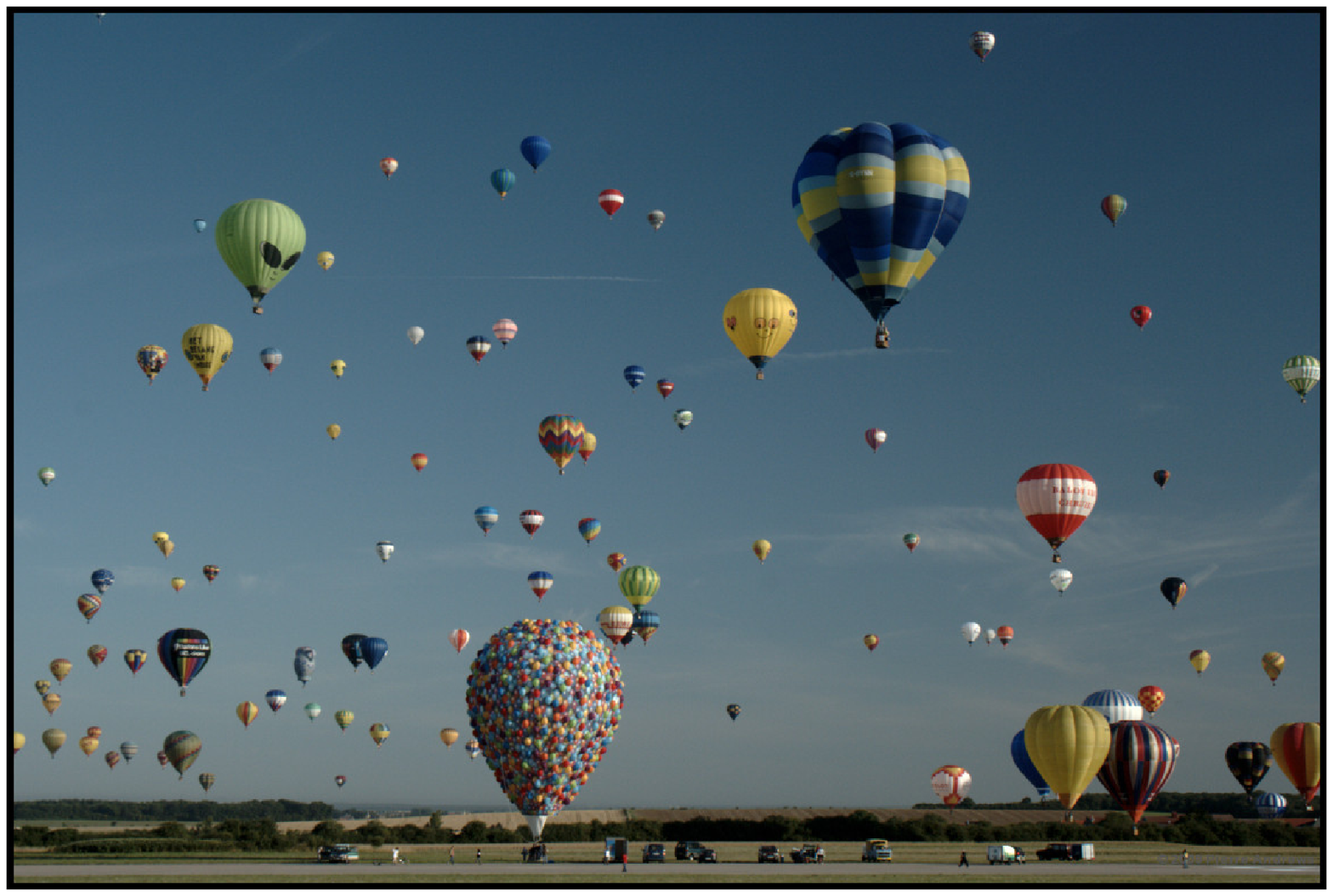,width=0.48in,height=0.36in}\epsfig{figure=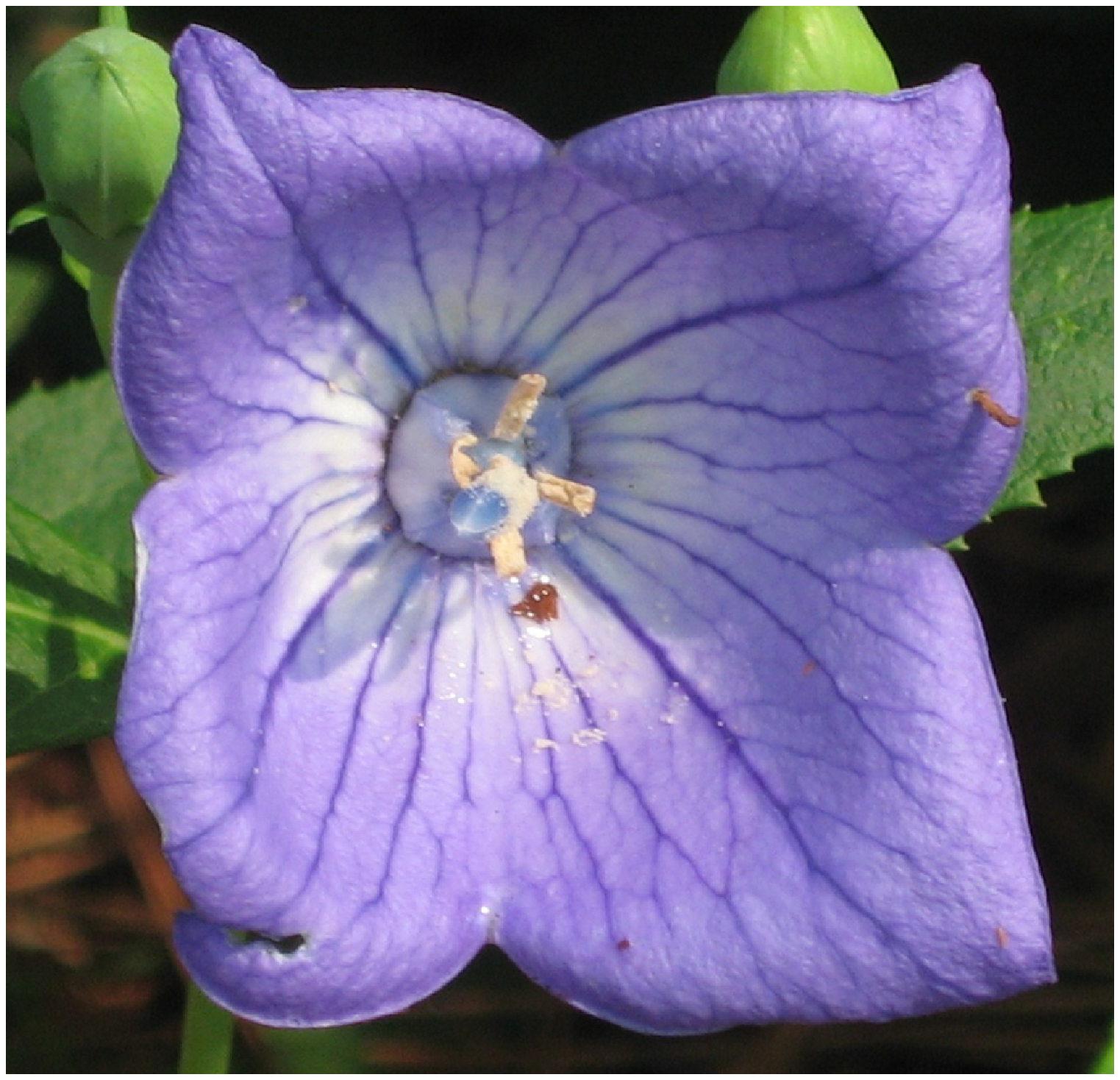,width=0.48in,height=0.36in}}
\subfigure[snow leopard] {\epsfig{figure=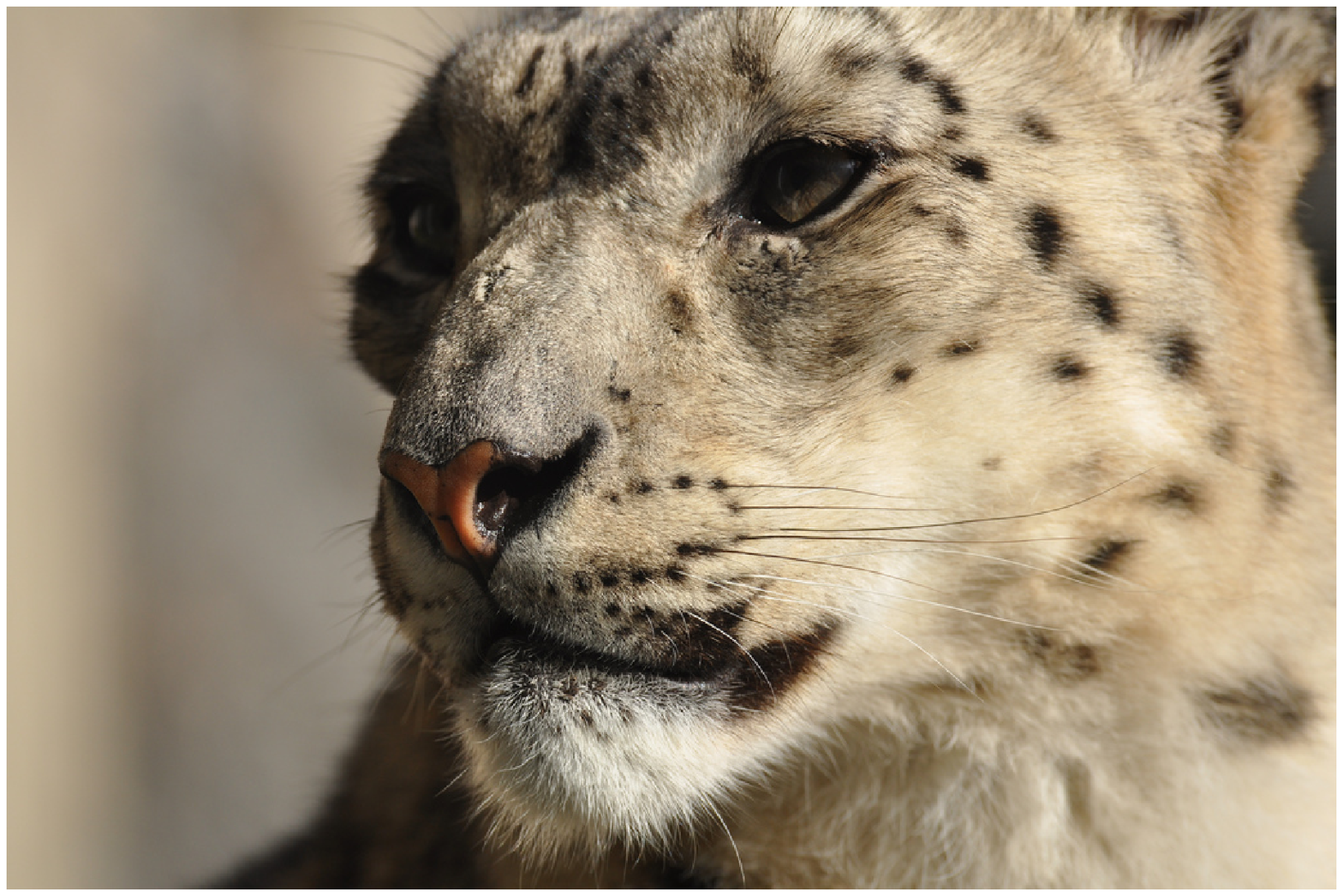,width=0.48in,height=0.36in}\epsfig{figure=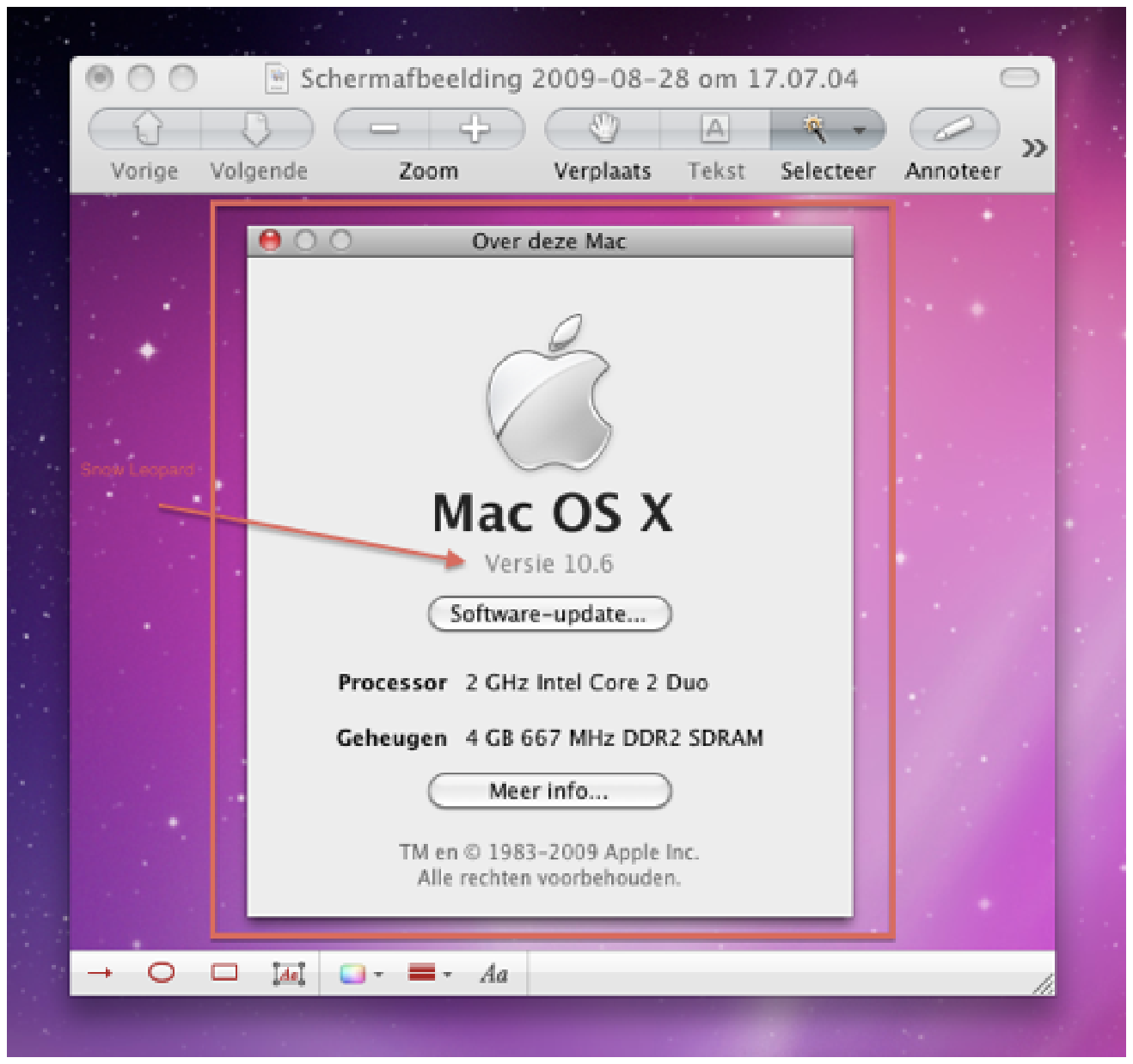,width=0.48in,height=0.36in}}\\
\subfigure[guitar] {\epsfig{figure=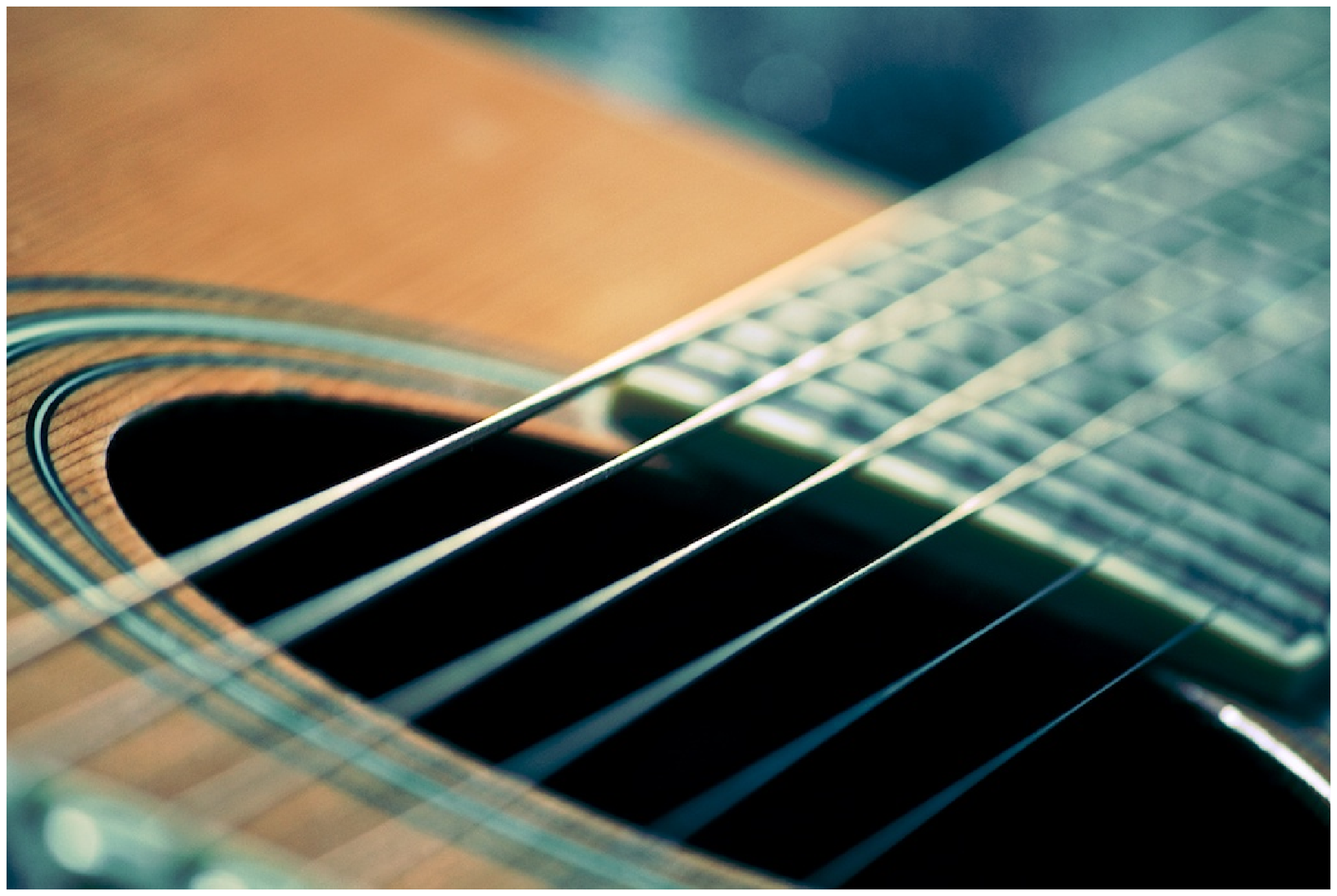,width=0.48in,height=0.36in}\epsfig{figure=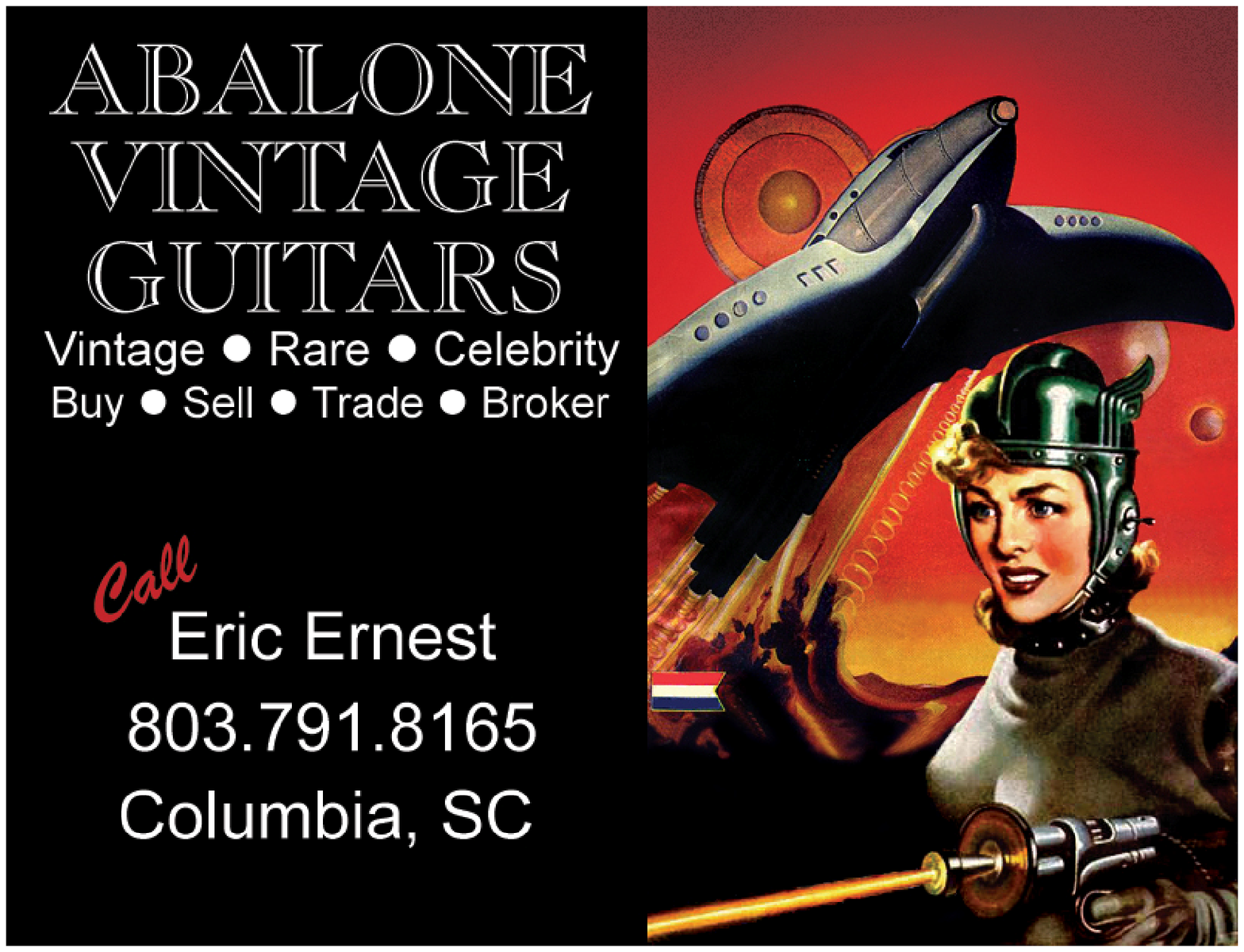,width=0.48in,height=0.36in}}
\subfigure[pumpkin] {\epsfig{figure=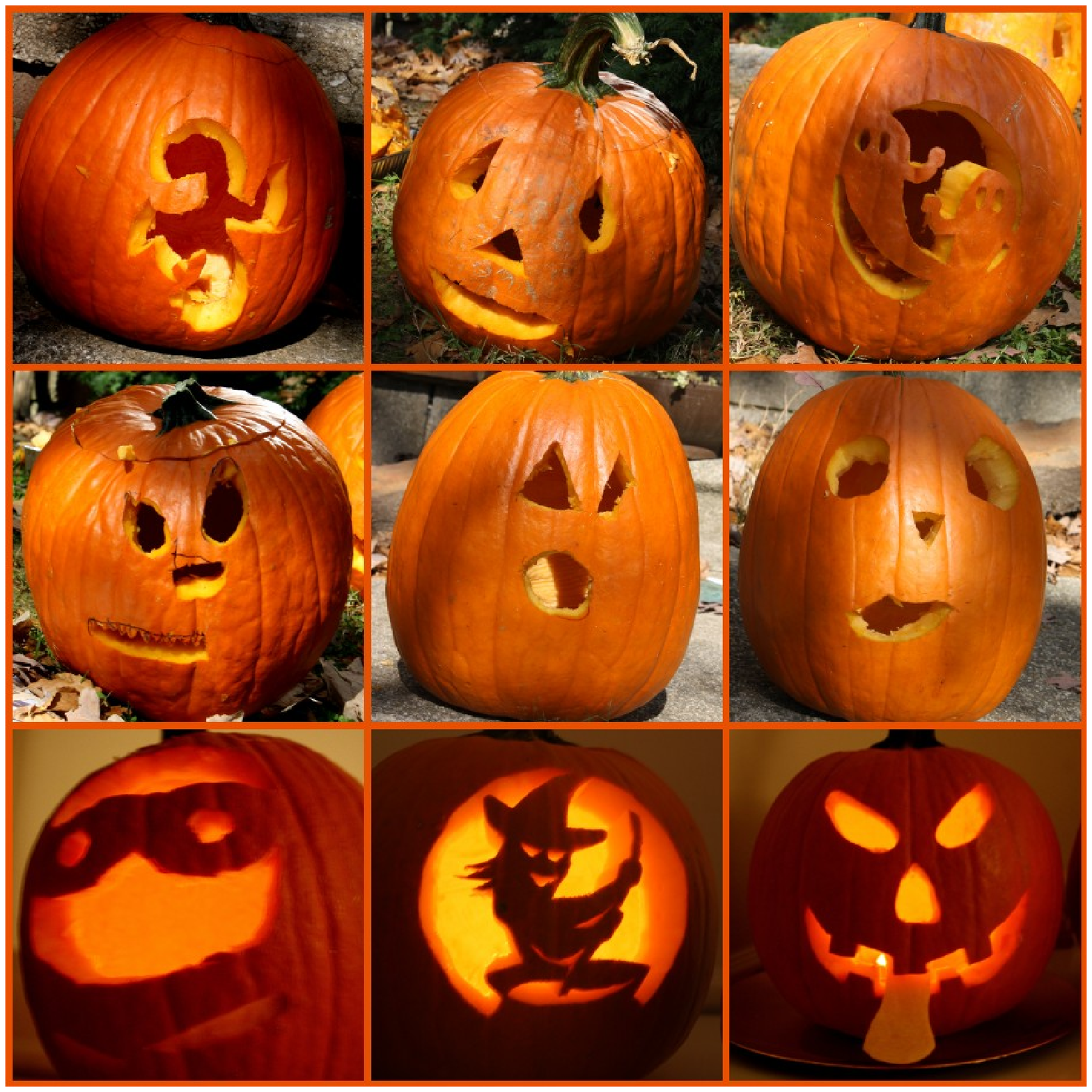,width=0.48in,height=0.36in}\epsfig{figure=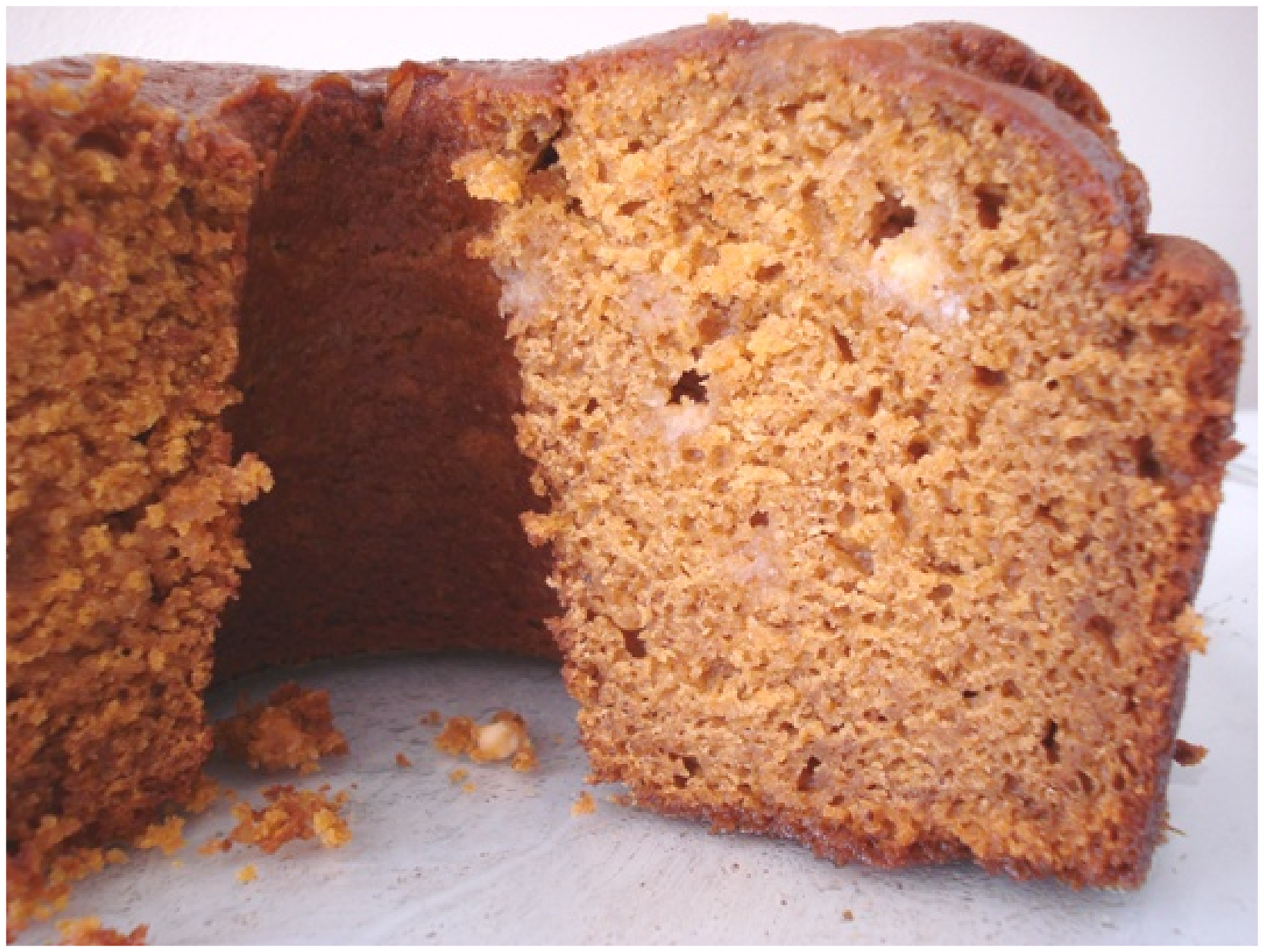,width=0.48in,height=0.36in}}
%\vspace*{-0.1in}
%}
%\mbox{
%\subfigure[balloon] {\epsfig{figure=Figures/pic/balloon.eps,width=0.56in,height=0.42in}\epsfig{figure=Figures/pic/balloon2.eps,width=0.56in,height=0.42in}}
%\subfigure[snow leopard] {\epsfig{figure=Figures/pic/snowleopard.eps,width=0.56in,height=0.42in}\epsfig{figure=Figures/pic/snowleopard2.eps,width=0.56in,height=0.42in}}
%\subfigure[guitar] {\epsfig{figure=Figures/pic/guitar.eps,width=0.56in,height=0.42in}\epsfig{figure=Figures/pic/guitar2.eps,width=0.56in,height=0.42in}}
%%\vspace*{-0.1in}
%}
\caption{Examples of image and the associated user tags in Flickr data set.  In each subfigure the left image  is correctly tagged by users, while the right one is wrongly tagged.}\label{Fig:Fig03}
\label{expmiss}}
\end{figure}

{\bf User tagging data set:} We also evaluate the algorithm on a
user tagging data set from an online image sharing web site {\em
Flickr.com}. This data set contains $13,528$ users (data sources)
who annotate $36,280$ images (data objects) with their own tags. We
consider $12$ tags - ``balloon," ``bird," ``box,"  ``car,"   ``cat,"
``child," ``dog," ``flower," ``snow leopard,"    ``waterfall,"
``guitar," ``pumpkin" for evaluation purposes. Each tag is
associated with a binary value 1/0 to represent its presence or not
in an image, and we apply MSS model to these $12$ tags separately to
find whether they are present on each image. To test  accuracy, we
manually annotate these $12$ tags on a subset of $1,816$ images.
%Different from the book author data set, each image has a $8,000$
%dimensional hierarchical gaussian \cite{Zhou:ICCV09} feature vector
%to represent its visual content. This imposes a more informative
%prior on the data value for each image.
Figure \ref{Fig:Fig03}
illustrates some image examples in this data set and the tags
annotated by users.
We can find some images are wrongly tagged by users.  The MSS model aims to
correct these errors and yield accurate annotations on these images.
%Actually, on this Flickr data set, most of images only have been annotated by one user, which means the annotated images by different users do not overlap much.

We follow the same experimental setup as on the book author data set.
%For the sake of fair
%comparison, we adopt the idea in \cite{Gupta:WWW11} to incorporate visual features to
%enhance the
%original algorithms for comparison which infers the true values based
%on object clusters in the feature space.  It has shown better
%accuracy compared with the  original algorithms \cite{Gupta:WWW11}.
Table \ref{Tb:Tb01} shows the average precision and recall on the 12
tags by the compared algorithms. We can see that {\em MSS} still
performs the best among these compared algorithms.

%In addition,
%since feature vectors are used to represent each image on the Flickr data set, we also compare MSS with the tag de-noising approach, which infers the true tags for each image with the low-rank prior as well as smoothness assumption that asserts visual similar images ought to have similar tags.  This low-rank approach involves the image feature vectors to compute the visual similarity between images.  It has achieved the state-of-the-art performance compared with the other compared algorithms in.  On this Flickr data set, the low-rank approach achieves an accuracy of $0.$.
%
%In addition to low-rank approach, we use the tags voted by users as baseline (VotedTag), as well as compare with MSSNoRank.  Table \ref{Tb:Tb03} shows the comparison results where MSS performs the best among the compared approaches.  This demonstrates with the feature vector information, MSS still outperforms the other approaches.

We also compare the computational time used by different algorithms
in Table \ref{Tb:Tb03}. The experiments are conducted on a personal
computer with Intel Core i7-2600 3.40 GHz CPU, 8 GB physical memory
and Windows 7 operating system. We can see that compared with most
of other algorithms, MSS model can converge in fewer rounds with
less computational cost.

\section{Conclusion}
In this paper, we propose
an integrated true
value inference and group reliability approach.
Dependent sources which are grouped together, and their (general and specific)
reliability is
assessed at group level.
The true data values are extracted from the reliable groups so that
the risk of overusing the observations from dependent
sources can be minimized.  The overall approach is described by a
probabilistic multi-source sensing model,
based on which we jointly infer
group reliability as well as the true values for objects {\em a posterior}
given the observations from sources.
The key to the
success of this model is
to capture the dependency between sources, and
aggregate the collective knowledge at the group granularity.
We present
experimental results  on two real data sets, which demonstrate the
effectiveness of the proposed model over other existing algorithms.

% Acknowledgements should only appear in the accepted version.
%\section*{Acknowledgments}

% In the unusual situation where you want a paper to appear in the
% references without citing it in the main text, use \nocite
%\nocite{langley00}

{
\bibliography{sigproc,truth_sigproc}

\begin{thebibliography}{11}
\providecommand{\natexlab}[1]{#1}
\providecommand{\url}[1]{\texttt{#1}}
\expandafter\ifx\csname urlstyle\endcsname\relax
  \providecommand{\doi}[1]{doi: #1}\else
  \providecommand{\doi}{doi: \begingroup \urlstyle{rm}\Url}\fi

\bibitem[Bachrach et~al.(2012)Bachrach, Minka, Guiver, and
  Graepel]{Bachrach:ICML12}
Bachrach, Y., Minka, T., Guiver, J., and Graepel, T.
\newblock How to grade a test without knowing the answers - a bayesian
  graphical model for adaptive crowdsourcing and aptitude testing.
\newblock In \emph{Proc. of International Conference on Machine Learning},
  2012.

\bibitem[Dong et~al.(2009)Dong, Berti-Equille, and Srivastava]{Dong:VLDB09b}
Dong, X.~L., Berti-Equille, L., and Srivastava, D.
\newblock Integrating conflicting data: The role of source dependence.
\newblock In \emph{Proc. of International Conference on Very Large Databases},
  August 2009.

\bibitem[Galland et~al.(2010)Galland, Abiteboul, Marian, and
  Senellart]{Galland:WSDM10}
Galland, A., Abiteboul, S., Marian, A., and Senellart, P.
\newblock Corroborating information from disagreeing views.
\newblock In \emph{Proc. of ACM International Conference on Web Search and Data
  Mining}, February 2010.

\bibitem[Jordan et~al.(1999)Jordan, Ghahramani, Jaakkola, and
  Saul]{Jordan:MLJ99}
Jordan, M., Ghahramani, Z., Jaakkola, T., and Saul, L.
\newblock Introduction to variational methods for graphical models.
\newblock \emph{Machine Learning}, 37:\penalty0 183--233, 1999.

\bibitem[Kasneci et~al.(2011)Kasneci, Gael, Stern, and Graepel]{Kasneci:WSDM11}
Kasneci, G., Gael, J.~V., Stern, D., and Graepel, T.
\newblock Cobayes: Bayesian knowledge corroboration with assessors of unknown
  areas of expertise.
\newblock In \emph{Proc. of ACM International Conference on Web Search and Data
  Mining}, 2011.

\bibitem[Kurihara et~al.(2006)Kurihara, Welling, and Vlassis]{Kurihara:NIPS06}
Kurihara, K., Welling, M., and Vlassis, N.
\newblock Accelerated variational dirichlet process mixtures.
\newblock In \emph{NIPS}, 2006.

\bibitem[Pasternack \& Roth(2010)Pasternack and Roth]{Pasternack:COLING10}
Pasternack, J. and Roth, D.
\newblock Knowing what to believe (when you already know something).
\newblock In \emph{Proc. of International Conference on Computational
  Linguistics}, August 2010.

\bibitem[Sethuraman(1994)]{Sethuraman:SS94}
Sethuraman, J.
\newblock A constructive definition of dirichlet priors.
\newblock \emph{Statistica Sinica}, 4:\penalty0 639--650, 1994.

\bibitem[Yin \& Tan(2011)Yin and Tan]{Yin:WWW11}
Yin, X. and Tan, W.
\newblock Semi-supervised truth discovery.
\newblock In \emph{Proc. of International World Wide Web Conference}, March
  28-April 1 2011.

\bibitem[Yin et~al.(2007)Yin, Han, and Yu]{Yin:KDD07}
Yin, X., Han, J., and Yu, P.~S.
\newblock Truth discovery with multiple conflicting information providers on
  the web.
\newblock In \emph{Proc. of ACM SIGKDD conference on Knowledge Discovery and
  Data Mining}, August 2007.

\bibitem[Zhao et~al.(2012)Zhao, Rubinstein, Gemmell, and Han]{Zhao:PVLDB12}
Zhao, B., Rubinstein, B. I.~P., Gemmell, J., and Han, J.
\newblock A bayesian approach to discovering truth from conflicting sources for
  data integration.
\newblock In \emph{Proc. of International Conference on Very Large Databases},
  2012.

\end{thebibliography}
\bibliographystyle{icml2013}
}
\end{document}